\documentclass[twocolumn,showpacs,preprintnumbers,amsmath,amssymb]{revtex4}

\usepackage{graphicx}
\usepackage{dcolumn}
\usepackage{bm}

\newcommand{\Fref}[1]{Figure~{\ref{#1}}}
\newcommand{\fref}[1]{Fig.~{\ref{#1}}}

\newcommand{\Tref}[1]{Table~{\ref{#1}}}
\newcommand{\Cref}[1]{Chapter~{\ref{#1}}}

\begin{document}

\title{Catastrophic Fermi surface reconstruction in the shape-memory alloy AuZn}
\author{P. A. Goddard}
\author{J. Singleton}
\author{R. D. McDonald}
\author{N. Harrison}
\author{J. C. Lashley}
\affiliation{National High Magnetic Field Laboratory, Los Alamos National Laboratory, MS-E536, Los Alamos, New Mexico 87544, USA}
\author{H. Harima}
\author{M.-T. Suzuki}
\affiliation{Physics Department, Kobe University, Kobe 657-8501, Japan}

\date{\today}

\begin{abstract}
AuZn undergoes a shape-memory transition at 67~K. The de Haas van Alphen effect persists to 100~K enabling the observation of a change in the quantum oscillation spectrum indicative of a catastrophic Fermi surface  reconstruction at the transition. Coexistence of both Fermi surfaces at low temperatures is suggestive of an intrinsic phase separation in the bulk of the material. In addition, a Dingle analysis reveals a sharp change in the scattering mechanism at a threshold cyclotron radius, which we suggest to be related to the underlying microstructure that drives the shape-memory effect.
\end{abstract}


\maketitle

A deformed wire that returns to its original shape on heating; metallic eyeglass frames with rubber-like flexibility; orthodontic arch-wires and brassiere supports that provide a constant force over long periods of time and significant changes in load; these are just a few of the applications of shape-memory alloys~\cite{otsuka}. 

More specifically, a shape-memory alloy (SMA) is one in which plastic strain is recovered on heating the material above the temperature at which a structural martensite to austenite phase transition takes place~\cite{otsuka,bhattacharya}. At temperatures just above the transition the stress-strain loop shows a hysteresis that leads to superelasticity~\cite{otsuka,bhattacharya}. The high-temperature austenite phase and the low-temperature martensite phase are so-named after the analogous phases in steel. However, steel, like many other martensitic materials, does not exhibit the shape-memory effect -- SMAs are a sub-set of materials with austenite-martensite phase transitions. 

Despite the technological significance of both the shape-memory effect and the martensitic transition (the transition in steel has been described as the world's most economically important metallurgical transformation~\cite{wayman}), not a great deal is known about the underlying microscopic mechanisms that drive them. However, given that all the systems that exhibit both these phenomena show metallic bonding it is likely that the electrons play a substantial role. In a previous study the SMA AuZn was shown to exhibit a significant change in magnetoresistance at its transition temperature, indicating an alteration of the Fermi surface associated with the transition~\cite{ross}. Citing the agreement between the observed low-temperature quantum oscillations and band-structure calculations together with the change in magnetoresistance as evidence, the authors of Reference~\cite{ross} suggest that it is a Fermi surface nesting event that drives the martensitic transition in this material. They further submit that the slight incommensurability of the proposed nesting vector causes a reversible symmetry reduction in the unit cell at the transition that in turn leads to the shape-memory effect. In this paper we offer conclusive proof of the Fermi surface reconstruction at the martensitic transition in AuZn by measuring de Haas-van Alphen (dHvA) oscillations above and below the transition temperature. 

\begin{figure}[b]
\centering
\includegraphics[height=6.9cm]{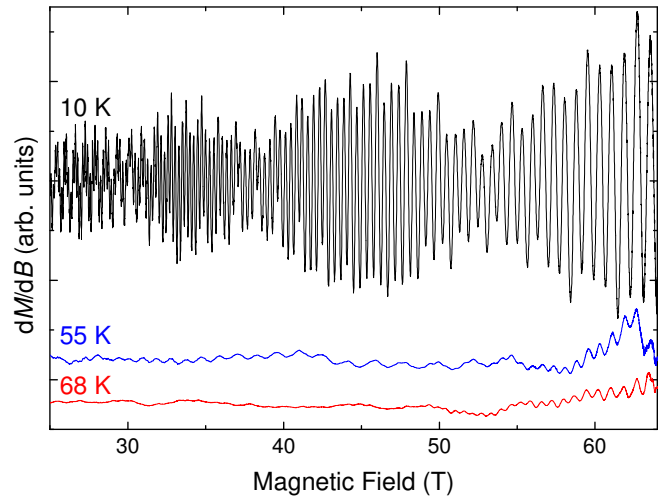}
\caption{De Haas-van Alphen oscillations observed in the magnetic susceptibility of AuZn at 10~K (upper), 55~K (middle) and 68~K (lower). The magnetic field is applied parallel to the [110] direction of the high temperature austenite phase. The curves are offset for clarity.} \label{dhva}
\end{figure}

\begin{figure*}[t]
\centering
\includegraphics[height=10cm]{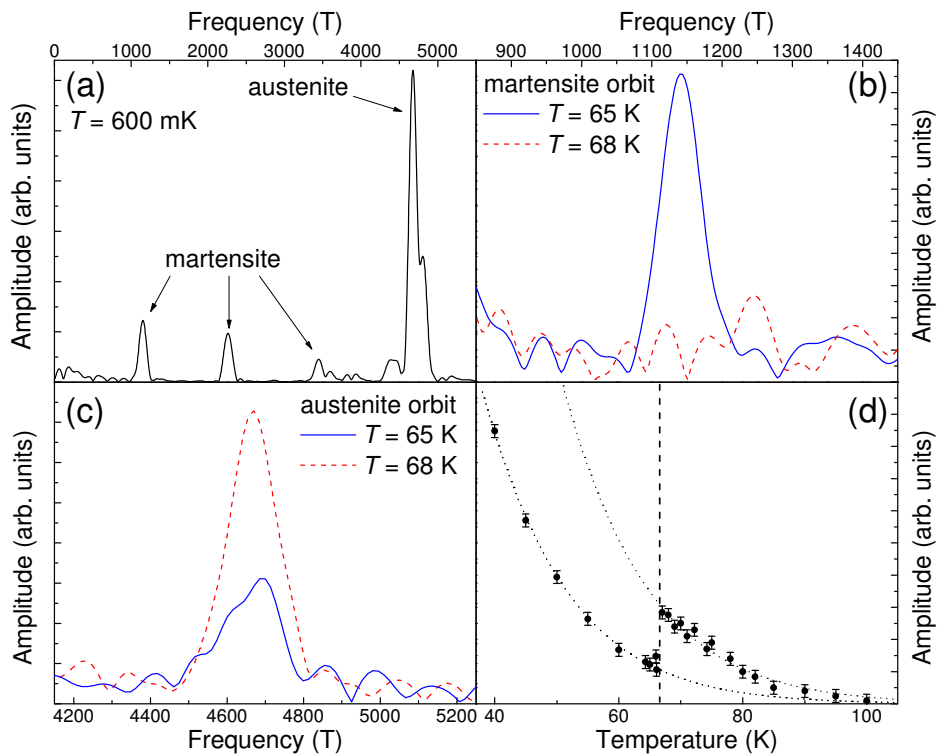}
\caption{(a) Fourier transform spectra of the de Haas-van Alphen oscillations at 600~mK. (b) Detail of the peak at 1140~T at two temperatures straddling the phase transition at $T=66.5$~K. (c) Detail of the peak at 4660~T at the same two temperatures. (d) The amplitudes of the 4660~T peak (data points) as a function of temperature. The sharp drop in the amplitude occurs at the phase transition. The dotted lines represent fits to the Lifshitz-Kosevich formula described in the text. All the data are taken with the field applied parallel to the austenite [110] direction.} \label{fft}
\end{figure*}

To allow successful observation of the dHvA oscillations in both phases, the sample must fulfill two criteria; first, it must be an exceptionally clean system, which in the case of alloys also implies stoichiometry; second, its transition temperature must be as low as possible. AuZn is thus ideal, having the lowest transition temperature of any stoichiometric SMA yet recorded~\cite{darling, lashley} and the added advantage of having no precursor phases prior to formation of the martensite~\cite{ross}. 

The transformation in AuZn is believed to take place as follows~\cite{makita, barsch}; as the material is cooled through the transformation temperature the cubic unit cell of the austenite phase distorts along the [110] direction. A strain and commensurate shuffle leads to the hexagonal primitive unit cell of the martensite phase forming along four possible orientations corresponding to the four body diagonals of the austenite. It is these four equally likely martensite variants that provide a macroscopic explanation for the shape-memory effect. Simply put, if a deformation of  the sample that takes place below the transition temperature can proceed via the inter-conversion of the martensite variants then the original shape will be recovered when all the variants transform back into the single crystal austenite on heating~\cite{bhattacharya}.

The dHvA measurements presented here are performed using a highly compensated coil magnetometer in pulsed magnetic fields of up to 65~T and at temperatures ranging from 600~mK to 100~K. The dHvA effect measured in this way is sensitive to the majority of the sample; thus, it is a bulk effect and yields information that cannot be provided by surface measurements, such as microscopy, alone. The fact the measurements are possible at 100~K, which appears to be the highest temperature at which dHvA oscillations have yet been observed, is due to the combination of high magnetic fields, a material uncommonly free of impurities, and quasiparticles with low effective masses.

\Fref{dhva}~shows the measured dHvA oscillations at three different temperatures with the magnetic field applied parallel to the [110] direction of the austenite phase. In this study the martensitic transition is found to occur at $66.5\pm0.5~$K, and hence the oscillations at 68~K must be produced by extremal quasiparticle orbits on the austenite Fermi surface. At temperatures below the transition other frequencies are seen that presumably arise from the martensite Fermi surface.
 
\Fref{fft}~(a) shows the Fourier-transform spectra of the data at 600~mK. Several distinct frequencies of dHvA  oscillation are observed. The frequency around 1140~T and its second and third harmonics are attributed to the martensite phase, while the frequency around 4660~T is attributed to the austenite phase. The justification for this attribution is illuminated in the rest of the figure. \Fref{fft}~(b) shows the 1140~T peak at two temperatures that straddle the phase transformation. This frequency is not present at 68~K, but only appears at temperatures below the transition, demonstrating that it is caused by orbits on the martensite Fermi surface. Conversely, \fref{fft}~(c) shows that the peak at 4660~T is present at 68~K implying that it is due to orbits on the austenite Fermi surface. It is seen that the amplitude of this frequency drops as the temperature is cooled below the transformation temperature. This is more clear in \fref{fft}~(d) which shows the temperature dependence of the amplitude of the 4660~T frequency. A sharp drop occurs at 66.5~K caused by a large part of the austenite transforming into the martensite phase and hence no longer being able to support this frequency of oscillation.

\begin{table} \caption{The measured frequencies, $F_{meas}$, and effective masses, $m^*_{meas}$, of de Haas-van Alphen oscillation in units of Tesla and the electronic rest mass, respectively. Also shown are the frequencies and effective masses from the band-structure calculations that best correspond to the measured values. These calculated orbits are shown in \fref{calcFS}.} 
\centering 
\begin{tabular}{ccccc} 
\hline 
phase & $F_{meas}$~(T) & $m^*_{meas}$ ($m_{\rm e}$) & $F_{calc}$~(T) & $m^*_{calc}$ ($m_{\rm e}$)\\ 
\hline 
martensite & $46\pm1$ & $0.051\pm0.002$ & 73 & 0.119\\
martensite & $303\pm1$ & $0.111\pm0.006$ & 588 & 0.117\\
martensite & $1141\pm1$ & $0.195\pm0.003$ & 1024 & 0.262\\
austenite & $4669\pm7$ & $0.35\pm0.01$ & 4598 & 0.346\\
\hline 
\label{compare} 
\end{tabular} 
\end{table}

 \begin{figure}[b]
\centering
\includegraphics[height=8cm]{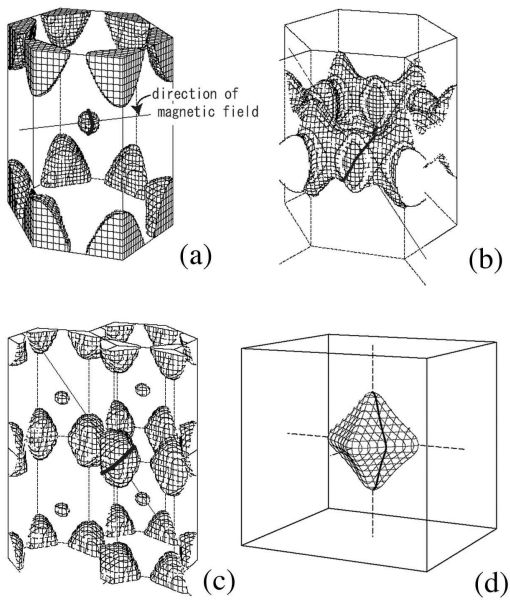}
\caption{Fermi surface sections of AuZn obtained from the band-structure calculations, together with the quasiparticle orbits (bold lines) that best correspond to the four measured de Haas-van Alphen oscillations. (a), (b), (c) and (d) occur in the order in which the frequencies appear in \Tref{compare}. There are four martensite variants corresponding to the four body diagonals of the cubic austenite phase~\cite{ross}. Thus when the magnetic field is applied parallel to the [110] direction of the austenite, geometry dictates that the field lies along two inequivalent directions in the martensite Brillouin zone. One direction leads to orbits whose plane is perpendicular to the hexagonal plane of the Brillouin zone, as is the case in \fref{calcFS}~(a), while the other direction leads to orbits whose plane is inclined at around $60^{\circ}$ to the hexagonal plane, as is the case in \fref{calcFS}~(b) and (c). \Fref{calcFS}~(d) shows the austenite Fermi surface and orbit.} \label{calcFS}
\end{figure}

Taken together \fref{fft}~(a), (b), (c) and (d) clearly show that there is a significant Fermi surface reconstruction associated with the martensitic phase transformation, and that the portion of the austenite phase that survives the transition coexists with the martensite down to very low temperatures.
 
The dotted lines in \fref{fft}~(d) are fits of the data to the temperature-dependent portion of the Lifshitz-Kosevich formula for amplitude of quantum oscillations, which has the form $A (\gamma T/B) / \sinh(\gamma T/B)$, where $A$ is proportional to the amount of the sample that contributes to the oscillation, $\gamma \approx 14.7 m^*$, $m^*$ is the quasiparticle effective mass in units of the electronic rest mass, $B$ is the magnetic field and $T$ is the temperature~\cite{shoenberg}. The discontinuity requires that two fits be made, one above the transition and one below, both with $A$ and $m^*$ as free parameters. The effective masses taken from the fits to the  high- and low-temperature data are found to be the same within the errors, the only difference between the fits arising from the $A$-parameters. The ratio of the high- and low-temperature values of $A$ gives the volume fraction of the austenite phase remaining below the transition, and is found to be $50\pm16\%$ for the data in \fref{fft}~(d). Several experiments have been performed and the exact volume fraction is found to vary, probably due to slight changes in the stress on the sample associated with the method with which it is fixed into the measurement coil. Whatever the reason, the volume fraction never drops below about 20\% even when the sample is completely free. The loss of austenite occurs abruptly at the transition temperature and there is no evidence of further reduction down to low temperatures. This is in agreement with other experimental observations in AuZn, such as dilatometry or resistance~\cite{ross}, which show a sharp transition with relatively little hysteresis. Thus it seems that a certain amount of austenite phase gets ``locked-in'' on completion of the transition, i.e. a bulk phase separation of the austenite and martensite at low temperatures is an intrinsic property of the material.

By performing a Fourier analysis of data taken at many different temperatures over several ranges of magnetic field a total of three independent frequencies of dHvA oscillation due to the martensite phase are identified as well as the austenite oscillation. Their frequencies and effective masses are tabulated in \Tref{compare}.  

Band-structure calculations are performed in both the high and low temperature phases. They include the effects of spin-orbit interactions by means of a full-potential linear augmented-plane wave (FLAPW) method. The exchange-correlation potential was modeled by means of a local-density approximation (LDA). The calculated orbits that best correspond to the measured frequencies, together with the Fermi surface sections on which they reside are shown in \fref{calcFS}. 

The calculated frequencies and effective masses are also shown in \Tref{compare}. Band-structure calculations have previously been successful at qualitatively predicting the angle-dependence of the quantum oscillations in the martensite phase of AuZn~\cite{ross}. However, \Tref{compare} highlights the quantitative discrepancies between the calculations and the measured values that arise from the quality of data pertaining to the low-temperature crystal structure, x-ray diffraction measurements being difficult due to the density of the material. Using neutron scattering the transformation was found to occur via a mechanism similar to that in AuCd~\cite{makita}. The martensite structure was then deduced using the established group-theoretical arguments for AuCd~\cite{barsch}. However, the exact size of the distortion occuring at the transition in AuZn is not known and any differences between this distortion and the equivalent in AuCd will lead to the discrepancies between the experimental and calculated results. That the agreement between the calculated and measured values of the austenite orbit is excellent reflects the simplicity of the high-temperature, cubic unit cell compared to the complexity of the modified structure after the phase transition has taken place.

\begin{figure}[t]
\centering
\includegraphics[height=6.9cm]{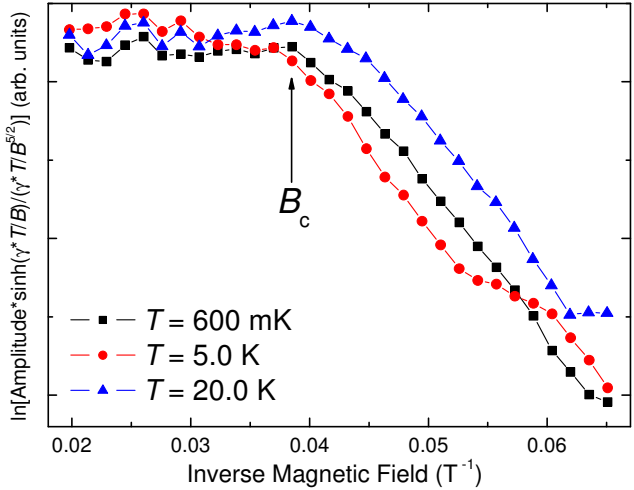}
\caption{Result of performing a Dingle analysis on the amplitudes of the 1141~T frequency of martensite dHvA oscillation at three temperatures. It is suggested that  $B_{\rm c}$ corresponds to the field at which the cyclotron radius ($\approx 50$~nm) is representative of the size of a single martensite variant within the bulk of the sample.} \label{ding}
\end{figure}

\Fref{ding} shows the result of performing a Dingle analysis of the 1141~T martensite frequency of dHvA oscillation using the three-dimensional Lifshitz-Kosevitch formula~\cite{shoenberg} at three different temperatures. Note that the abscissa is in inverse magnetic field units. The plateau at high values of the magnetic field represents a region where the scattering rate takes a constant value~\cite{shoenberg}. At the point marked with an arrow the plateau-like region ends and as the magnetic field is lowered the scattering rate increases. We suggest that this point corresponds to the cyclotron radius becoming larger than a certain value representative of the martensitic microstructure. The cyclotron radius, $l_{\rm c}$, is the characteristic size of the orbitally-quantized wavefunction and is given by $l_{\rm c}=\sqrt{(2l_{\rm LL}+1)\hbar/eB}$, where $l_{\rm LL}=B_{\rm F}/B$ is the Landau level index and $B_{\rm F}$ is the fundamental frequency of quantum oscillation in Tesla. A tentative explanation for the data of \fref{ding} is that when the magnetic field is large the cyclotron radius is small and the amplitudes of the dHvA follow the Lifshitz-Kosevitch formula with a constant scattering rate determined by impurities within the martensite variant. As the field is lowered the cyclotron radius grows to the typical size of the martensite variant domains and increasing numbers of quasiparticles encounter the boundaries between the domains causing the scattering rate to increase. From the data we find that $B_{\rm c}$ corresponds to a cyclotron radius of approximately 50~nm.

In summary, a significant Fermi surface reconstruction has been directly observed via the quantum oscillations at  a structural phase transformation for the first time. In the specific case of AuZn the transition is shape-memory and an apparently intrinsic, bulk separation of the austenite and martensite phases is shown to exist at low temperatures, implying that due to the nature of the distortion a complete transformation is not possible for this material. A Dingle analysis of the oscillations suggests that characteristic size of the martensitic microstructure is around 50~nm in the bulk of the sample, an observation hitherto impossible using only surface techniques.

The authors would like to thank Peter Littlewood and Walt Harrison for stimulating discussions. This work is supported by the U.S. Department of Energy (DOE) under Grant No. LDRD-DR 20030084. Part of this work was carried out at the National High Magnetic Field Laboratory, which is supported by the National Science Foundation, the State of Florida and DOE.


\end{document}